\documentclass[12pt]{article}
\usepackage{amsmath,amssymb,graphicx,mathrsfs,hyperref}

\newcommand{\be}{\begin{equation}}
\newcommand{\ee}{\end{equation}}
\newcommand{\bea}{\begin{eqnarray}}
\newcommand{\eea}{\end{eqnarray}}

\def\({\left(} \def\){\right)}

\begin{document}
%%%%%%%%%%%%%%%%%%%%%%%%%%%%%%%%%%%%%%%%%

\title{\vspace{-1.8in} \begin{flushright} {\footnotesize CERN-PH-TH/2011-137,LMU-ASC 11/12}  \end{flushright}
\vspace{3mm}
\vspace{0.3cm} Wave function of the quantum black hole}
\author{\large Ram Brustein ${}^{(1,2)}$, Merav Hadad ${}^{(1,3)}$ \\ \vbox{\begin{flushleft} {\small
(1) Department of Physics, Ben-Gurion University,
    Beer-Sheva 84105, Israel  } \\ \vspace{7pt}   {\small
(2) CAS, Ludwig-Maximilians-Universit�at M\"unchen, 80333 M\"unchen, Germany}\vspace{7pt} \\  {\small
(3)    Department of Natural Sciences, The Open University of Israel }\vspace{4pt} \\  {\small
 \hspace{.2in} P.O.B. 808, Raanana 43107, Israel}\vspace{7pt} \\  {\small
  \vspace{7pt} {\small \textrm
    E-mail: ramyb@bgu.ac.il,\ meravha@openu.ac.il} } \end{flushleft}}
 }
\date{}
\maketitle
%%%
\vspace{-0.5in}
\begin{abstract}
%\abstract{

We show that the Wald Noether charge entropy is canonically conjugate to the opening angle at the horizon. Using this canonical relation we extend the Wheeler-DeWitt equation to a Schr\"{o}dinger equation in the opening angle, following Carlip and Teitelboim. We solve the equation in the semiclassical approximation by using the correspondence principle and find that the solutions are minimal uncertainty wavefunctions with a continuous spectrum for the entropy and therefore also of the area of the black hole horizon. The fact that the opening angle fluctuates away from its classical value of $2 \pi$ indicates that the quantum black hole is a superposition of horizonless states. The classical geometry with a horizon serves only to evaluate quantum expectation values in the strict classical limit.
%}
\end{abstract}

%\keywords{Equation of State, field equation, Entropy}

%\preprint{}

%\begin{document}
%\maketitle
\newpage

Quantum black holes have attracted continued interest over decades since the discovery of their unique thermodynamics \cite{bek,Hawking}. The quantum mechanical nature of the black hole (BH) has been investigated in the semiclassical approximation starting with Bekenstein's proposal that the horizon area of the quantum BH should be quantized, further arguing that the spectrum of the horizon area should be evenly spaced \cite{beknc,mukhanov,bekmuk}.

The Wheeler-DeWitt (WDW) equation \cite{WDW}, the standard equation that is used to study the quantum mechanics of BH's, has been extended by Carlip and Teitelboim \cite{carlip} to a Schr\"odinger-like equation for the BH. Their reasoning was based on the canonical structure of Einstein gravity, making the key observation that the BH horizon should be considered as a boundary of spacetime  in addition to the boundary at infinity.  Therefore, they argued,  the opening angle at the horizon (to be defined precisely later) and its conjugate variable, the area of the horizon, should also appear in the equation in addition to the mass of the BH and its conjugate variable, the time separation at infinity.

In a parallel and contemporary development
Wald found, by studying the canonical structure of generalized theories of gravity for BH's with bifurcating Killing horizons, that the Noether charge of diffeomorphisms contains two contributions. One contribution  from infinity  could be identified as the mass of the BH and another contribution from the horizon which Wald identified as the entropy of the BH \cite{wald1,wald2}.

We combine the Wald definition of the entropy with Carlip and Teitlboim's treatment of the quantum BH.  We first show that the opening angle of the horizon is canonically conjugate to the Wald Noether charge entropy. We identify the relationship between the Lie derivative of the area of a $D-2$ hypersurface embedded in a $D$ dimensional spacetime and the extrinsic curvature of the surface. Next we use Brown`s methods \cite{brown} as applied in \cite{ramy+dan,ramy+joey} to show that the Lie derivative of the area of the $D-2$ hypersurface is canonically conjugate to the Wald entropy. Then, by showing that the Lie derivative of the area of the $D-2$ hypersurface is equal to the opening angle at the horizon, we show that the opening angle at the horizon is canonically conjugate to the Wald entropy.

Having found that the Wald entropy is canonically conjugate to the opening angle at the horizon we follow Carlip and Teitelboim and extend the WDW equation to  a Schr\"{o}dinger-like equation. Our equation is for the opening angle and the Wald entropy rather than the original form which was for the opening angle and the area of the horizon. That the area should be replaced by the Wald entropy in this context was first proposed by Medved \cite{joey}.

To solve the equation in the semiclassical approximation we rely on the correspondence principle. We require that the expectation value of the opening angle equals to $2\pi$ (see below) -- its classical value and that the expectation value of the entropy be equal to the a quarter of the area of the horizon in units of the effective coupling (see below) -- the classical value of the Wald Noether charge entropy. Further, we require that the average of the  entropy fluctuations  be equal to the classical value which is determined by the specific heat of the BH. To implement this last step we need to place the BH in  Anti deSitter (AdS) space to make its specific heat positive. This would be equivalent to the less physical situation of putting the BH in a box in Minkowski space. We further argue that since the entropy of a large BH is large, then by the central limit theorem, the probability distribution should be approximately Gaussian. This then fixes the form of the wavefunction in the entropy representation and allows us to calculate the wavefunction in the opening angle representation.

We find that the spectrum of the entropy is continuous in the semiclassical approximation. Then it follows that the continuous spectrum of $S_W$ implies a  continuous spectrum for the horizon area $A_H$. This is in contrast to the Bekenstein conjecture \cite{beknc,mukhanov,bekmuk}.

Our analysis has several advantages. We start from first principles and use the standard rules of quantum mechanics. Since we use the Wald formalism which applies to generalized theories of gravity, our results are applicable to such theories and not only to Einstein theory of gravity.

We begin by foliating the BH spacetime with respect to a space-like coordinate, (constant $r$ hypersurfaces, for example) following the procedure introduced in \cite{brown} and applied in \cite{ramy+dan,ramy+joey}. Here we only sketch the derivation, further details and precise definitions can be found in \cite{brown,ramy+dan}.

The unit normal to the hypersurface $\Sigma_{D-1}$ is $u_a$ and the hypersurface metric is given by $g_{ab} =h_{\alpha\beta}e_{\ a}^\alpha e_{\ b}^\beta-u_a u_b$ where $e_{\ a}^\alpha$ is a basis of tangent vectors to the hypersurface. Greek indices denote the induced coordinates on the
hypersurface. We single out a $D-2$ cross section of the hypersurface $\Sigma_{D-2}$ whose area is given by
\begin{eqnarray}
\label{D-2 surface}
A_{D-2}=-\frac{1}{2}\int\limits_{\Sigma_{D-2}}\hat{\epsilon}^{\alpha b}\epsilon_{\alpha b}\,.
\end{eqnarray}
Here $\epsilon^{cd}$ is a $D-2$ volume form given by $\epsilon^{cd} = \hat{\epsilon}^{cd}\bar{\epsilon}$ , $\bar{\epsilon}$ being the area element. The bi-normal vector to the area element $\hat{\epsilon}^{cd}=\nabla^c u^d$ is  normalized as $\hat{\epsilon}^{cd}\hat{\epsilon}_{cd}=-2$.

The area of the cross section $\Sigma_{D-2}$ and the Lie derivative of the area along $u_a$ $\mathcal{L}_u$, can both be expressed in terms of the hypersurface metric
\begin{eqnarray}
\label{D-2 surfaceh}
A_{D-2}&=&-\frac{1}{2}\int\limits_{\Sigma_{D-2}} g^{\alpha \gamma}g^{bd} \epsilon_{\gamma d} \epsilon_{\alpha b}=\int h^{\alpha \gamma}u^b u^d \hat{\epsilon}_{\gamma d}\epsilon_{\alpha b}, \\
\mathcal{L}_u A_{D-2}&=&-2\int\limits_{\Sigma_{D-2}} K^{\alpha \gamma}u^b u^d \hat{\epsilon}_{\gamma d}\epsilon_{\alpha  b},
\label{Lie D-2 surfaceh}
\end{eqnarray}
where $K^{\alpha \gamma}=-\frac{1}{2}\mathcal{L}_u h^{\alpha \gamma}$ is the extrinsic curvature and we have used the fact that both $\mathcal{L}_u u_a$ and $\mathcal{L}_u \hat{\epsilon}_{\alpha b}$ vanish.

The next step is to find the canonically conjugate variable to $\mathcal{L}_u A_{D-2}$. For this purpose we may use Brown`s results \cite{brown}  showing that the gravitational action contains the relevant term
$
\int d^Dx\sqrt{-g}\frac{\partial\mathscr{L}}{\partial R_{p\alpha\beta q}}u_p u_q \mathcal{L}_u K_{\alpha\beta}\, .
$
Then, the extrinsic curvature is the canonically conjugate variable to $\frac{\partial\mathscr{L}}{\partial R_{p\alpha\beta q}}u_p u_q$. Since $\mathcal{L}_u u_a=0$ we may rewrite this term as
\begin{eqnarray}
\label{action part1}
\int d^Dx\sqrt{-g}\frac{\partial\mathscr{L}}{\partial R_{p\alpha\beta q}} \mathcal{L}_u \left(K_{\alpha\beta}u_p u_q\right)
\end{eqnarray}
and conclude that $K_{\alpha\beta}u_p u_q$ is canonically conjugate to $\frac{\partial\mathscr{L}}{\partial R_{p\alpha\beta q}}$.  The projections of these variables on the bi-normal $\hat{\epsilon}^{\gamma d}$ satisfy standard Poisson bracket relations:
\begin{eqnarray}
\label{canonical}
\left\{K_{\alpha \gamma}u_b u_d \hat{\epsilon}_{\gamma d}\hat{\epsilon}_{\alpha  b}(x_1), \frac{\partial\mathscr{L}}{\partial R_{ b \alpha \gamma d}}\hat{\epsilon}_{\gamma d}\hat{\epsilon}_{\alpha  b}(x_2)\right\}=(-h)^{-1/2}\delta^{D-1}(x_1\!-\!x_2).
\end{eqnarray}

We wish to find the Poisson bracket between $\mathcal{L}_u A_{D-2}$ and the Wald Noether charge entropy,
\begin{equation}
S_W=-2\pi\oint\limits_{H} \frac{\partial\mathscr{L}}{\partial R_{\alpha b  \gamma d}}\hat{\epsilon}_{\gamma d}\epsilon_{\alpha  b}.
\label{walddef}
\end{equation}
Here we have used units in which the BH temperature is equal to $2\pi$. This choice of units will become relevant later.

To find the Poisson brackets we need to integrate Eq.~(\ref{canonical}) over the closed bifurcation surface of the BH horizon.  We need to perform a limiting procedure on a ``stretched horizon" because the BH horizon is a null surface while the $D-2$ hypersurface that we have discussed so far is time-like. In this limit the normal $u_a$ becomes the horizon Killing vector and the surface $\Sigma_{D-2}$ becomes the horizon bifurcation surface.  This limit will be taken at the end of the calculation, however, it is implicitly assumed in all of our calculations.

We now need to perform a double integral $\oint\limits_{\Sigma_{D-2}}\bar{\epsilon}(x_1) \oint\limits_{\Sigma_{D-2}} \bar{\epsilon}(x_2)$ over Eq.~(\ref{canonical}) and take the limit that the $D-2$ cross section becomes the bifurcation surface $\Sigma_{D-2}\to H$. Using Eqs.~(\ref{Lie D-2 surfaceh}) and (\ref{walddef}) we find that
\begin{eqnarray}
\label{canonical 2}
\left\{-\frac{1}{2}\frac{\mathcal{L}_u A_{D-2}}{ A_{D-2}}(t_1) , \frac{1}{2\pi}\, S_W(t_2)\right\}=(-g_{00})^{-1/2}\delta(t_1-t_2)
\end{eqnarray}
where we used the fact that $\oint\limits_{\Sigma_{D-2}}\!\!\!\bar{\epsilon}(x)=A_{D-2}$.

We may use the vanishing  of the Lie derivative of the $D-1$ volume $V^{D-1}$ of the foliated spacetime:  $\mathcal{L}_u V^{D-1}=0$. Since for a stationary spacetime the volume is a product of the proper time $\tau=\int\sqrt{-g_{00}}dt$ and the area $V^{D-1}= \tau A_{D-2}$ we find that $ A_{D-2}\mathcal{L}_u\tau +\tau \mathcal{L}_u A_{D-2}=0$ and thus we can express  $\mathcal{L}_u A_{D-2}$ in terms of $\Theta\equiv\frac{1}{2}\mathcal{L}_u\tau$, where $\Theta$ is the opening angle at the horizon. \cite{carlip} So $\Theta=-\tau \frac{1}{2} \frac{\mathcal{L}_u A_{D-2}}{A_{D-2}}$.

Integrating Eq.~(\ref{canonical 2})  over $\int d\tau =\int\sqrt{-g_{00}}dt_{1}$ and taking the limit that the $D-2$ cross section becomes the bifurcation surface $\Sigma_{D-2}\to H$ and $u$ becomes the horizon Killing vector we find, using eq.~(\ref{walddef}) that the  Poisson bracket between the Wald entropy and the opening angle at the horizon are given by
\begin{eqnarray}
\label{canonical 3}
\left\{\Theta,\frac{1}{2\pi}\,  S_W \right\}=1.
\end{eqnarray}

Identifying the Poisson bracket (\ref{canonical 3}) between the Wald entropy and the opening angle at the horizon allows us, following Carlip and Teitelboim \cite{carlip}, to canonically quantize the BH in the semiclassical approximation by extending the WDW equation to a  Schr\"{o}dinger-like equation. The wave function of the BH $\Psi$ then depends on $\Theta$ in addition to the time separation at infinity $T$ which is conjugate to the mass and to the other coordinates that are conjugate to conserved charges,
\begin{eqnarray}
\label{schrodinger1}
\frac{\hbar}{i}\delta\Psi(T,\Theta,\dots)+\left[ \delta t N^\mu H_\mu - \delta\Theta\, \frac{1}{2\pi} S_W+\delta T M+\cdots\right]\Psi(T,\Theta,\dots)=0. \end{eqnarray}
Here $M$ is the ADM mass, $H_\mu$ is the WDW Hamiltonian and the dots denote terms involving other conserved charges, such as the angular momentum, in case they are relevant. We refer to \cite{carlip} for the precise definitions of these quantities. The mass $M$ and the Noether charge entropy $S_W$ originate in a similar way as boundary integrals, one at infinity and the other on the horizon \cite{wald1}. They are related by the conservation of the Wald Noether charge via the first law of BH mechanics. From this point of view either $S_W$ or $M$ as well as all the other quantities appearing in Eq.~(\ref{schrodinger1}) are purely (quantum) mechanical quantities and do not have any statistical interpretation.

We find that it is more convenient to solve the equation in the entropy representation. Let us concentrate on the dependence of the BH wavefunction on $S_W$ which is governed by the equation
\begin{eqnarray}
\label{schrodinger}
\frac{2\pi \hbar}{i}\frac{\partial\Psi}{\partial S_W}=\Theta\Psi(S_W).
\end{eqnarray}

To solve Eq.~(\ref{schrodinger}) we need some  relation between $\Theta$ and $S_W$, $\Theta(S_W)$. We use the geometric and thermodynamic properties of large BH's to find such relations.

For a classical BH geometry $\Theta=2\pi$, reflecting the periodicity of Euclidean time. In Euclidean coordinates, angular nature and the geometric interpretation of the opening angle $\Theta$ is clearer and explicit. In Euclidean time $t_E=it$ the  proper time is given by $\tau_E=i\tau$ and the Euclidean opening angle is given by $\Theta_E=i\Theta$. In Euclidean Schwartzschild coordinates,
$
\Theta_E =\frac{1}{2} \frac{\sqrt{g_{00}}}{\sqrt{g_{11}}} \frac{ d t_E}{d r}
$.
Here  $dr$  denotes the coordinate distance away from the Euclidean origin. With the standard relation between the Euclidean time periodicity and the surface gravity at the horizon one can see that for a classical geometry with a regular Euclidean section $\Theta_E$ changes by $2\pi$ as a small circle around the origin is traversed.

So, we require that the expectation value of $\Theta$ be equal to $2\pi$, its classical value. We cannot impose the strict equality $\Theta=2\pi$ because then $S_W$ would be completely undetermined. We also cannot impose periodicity on $\Theta$. This would mean that classical geometries with values of $\Theta$ that differ by $2\pi$ would be equivalent. Clearly, a classical geometry with such a value of $\Theta$, for example $4\pi$, is singular and is not equivalent to a non-singular geometry with $\Theta=2\pi$. Had we imposed a periodic boundary condition on $\Theta$ we would have obtained a discrete evenly-spaced spectrum for the BH area. As we have just argued, this requirement is inconsistent with the semiclassical geometric point of view.

The geometrical interpretation of $S_W$ and its relation to the area of the BH horizon $A_H$ was found in \cite{waldarea}, where it was shown that
$
 S_W=\frac{A_H}{4 \hbar G_{eff}}
$
with a suitably defined effective coupling $G_{eff}$. In Einstein gravity $G_{eff}$ is equal to Newton's constant. In units in which the speed of light $c$ and Boltzmann constant $k$ are set to unity the effective coupling can be expressed in terms of an effective ``unit of area" which we define as $ A_{eff}\equiv\hbar G_{eff}$. In 4D Einstein gravity $A_{eff}$ is equal to the square of the Planck length.

So, we demand that the expectation value of $S_W$ be equal to its classical geometric value $\langle S_W\rangle=\frac{1}{4}\frac{A_H}{\hbar G_{eff}}$. Again, we cannot demand strict equality because then the value of $\Theta$ would be completely undetermined which is incompatible with the semiclassical geometric picture.

It follows that the resulting form of Eq.~(\ref{schrodinger}) is
\begin{eqnarray}
\label{schrodinger2}
\frac{\partial\Psi}{\partial S_W}=\Biggl[
\frac{2\pi i}{\hbar}-
a_1 \biggl(S_W-\langle S_W\rangle\biggr)+\cdots\Biggr] \Psi(S_W),
\end{eqnarray}
where the dots denote higher order in the Taylor expansion of $\Theta(S_W)$. The semiclassical BH has a very large entropy. This means that the fluctuations about the equilibrium point are approximately Gaussian with suppressed higher moments.
Then, the solution of Eq.~(\ref{schrodinger2}) with our imposed correspondence relations is
\begin{eqnarray}
\label{solution}
\Psi(S_W)= {\cal N}\
e^{-\frac{1}{2} a_1 \bigl(S_W-\langle S_W\rangle\bigr)^2}
e^{\ \hbox{\small $-\frac{2\pi i}{\hbar} S_W$}},
\end{eqnarray}
where ${\cal N}$ is a normalization factor.

To evaluate $a_1$ we use the correspondence principle again and argue that the quantum average of the entropy fluctuations $\Delta S_W ^2=\langle (S_W- \langle S_W\rangle)^2 \rangle$ should be equal to the classical value which is determined by the specific heat of the BH, $\Delta S_W ^2=C$. Since   $|\Psi|^2\sim Exp\left[{\hbox{-$\frac{(S_W-\langle S_W\rangle)^2}{2\Delta S_W^2}$}}\right]$ we find that $a_1=\frac{1}{2C} $ and  the wave function in the entropy representation is completely determined,
\begin{eqnarray}
\label{solutionfin}
\Psi(S_W)= {\cal N}\ e^{\hbox{$-\frac{\bigl(S_W-\langle S_W\rangle\bigr)^2}{4 C}$}} e^{\hbox{$-\frac{2\pi i}{\hbar} S_W$}}.
\end{eqnarray}

Having found the wave function in the entropy representation we can use the canonical commutation relations between $S_W$ and $\Theta$ to ``Fourier transform" it to the $\Theta$ representation,
\begin{eqnarray}
\label{solutiontheta}
\Psi(\Theta)= {\cal M}\ e^{\hbox{$-C\left(\Theta-2\pi\right)^2$}} e^{\hbox{\small $ \frac{i}{\hbar} \langle S_W\rangle \Theta$}},
\end{eqnarray}
where ${\cal M}$ is a normalization factor. As expected for Gaussian minimal uncertainty wave functions, $ \Delta S_W \Delta\Theta= \hbar/2$. 

In general, we expect $C$  for large BH's to be of the order $\langle S_W \rangle$, a very large number. This means that relative amplitude of the fluctuation in $S_W$ is very small. Similarly, The relative amplitude of the fluctuations in $\Theta$ away from $2\pi$ are very small.

As an example we consider Einstein Theory in 4D: In this case in order to calculate the specific heat we have to consider stable BH's with a positive specific heat. The most straightforward way is to consider BH's in AdS space \cite{HawkingPage}. The only difference between a BH in AdS space and in Minkowski space is a single negative mode so we expect the following results to apply also for the Minkowski space case if we neglect the unstable mode. The results of \cite{HawkingPage} imply that the specific heat of a large BH  is $C = 2 S$. In this case $C\sim A_H/l_p^2$, $l_p$ being the Planck length.

The wave function in Eq.~(\ref{solutiontheta}) can be though of as a superposition of wavefunctions with fixed $\Theta$. The deviation from the average value of $2\pi$ is very small for a large BH. However, no matter how small the deviation from $2\pi$ it has a crucial significance for the associated geometry. All geometries with $\Theta \ne 2\pi$ have a conical singularity and consequently no horizon. So we have found that the quantum BH, close as it may be to its classical counterpart, is actually a superposition of horizonless geometries. The classical geometry which does have a horizon serves only as a tool to evaluate quantum expectation values in the strict classical limit when $M_{BH}\to \infty$ and $G_{eff}\to 0$. This is perhaps related to the proposals in the context of string theory about the relationship between conical defect geometries and microstates of the BH \cite{vijayreview}.

Let us discuss now the properties of the spectrum of $S_W$.

The first observation is that the spectrum of $S_W$ is continuous. We emphasize that this does not necessarily imply a specific spectrum for the mass $M$ of a BH. In fact, in general, and independently of the area spectrum $M$  has a continuous spectrum. The most direct way of understanding this is to note that $T$, the counterpart of $\Theta$ at infinity (See Eq.~(\ref{schrodinger1})), does not have to be periodic or compact.  Then the spectrum of the conjugate variable $M$ is continuous.  The conservation law that they must obey (or equivalently, to satisfy  the first law) only requires that $\delta \left(\frac{\kappa}{2\pi}  \langle S_W \rangle\right)= \delta \langle  M\rangle$.

The second observation is that states with a large value of $S_W$ are highly degenerate. The most direct way to get to this conclusion is to take into account, as Carlip did \cite{carlip2}, that the algebra of diffeomorphisms on the horizon contains a Virasoro subalgebra.  If the central charge of this subalgebra and its energy are known, then the degeneracy of states for large values of the central charge and energy is given by the Cardy formula. The degeneracy of a state with eigenvalue $S_W$ is equal to $S_W$, so if one defines the entropy as the logarithm of the degeneracy of the state it is exactly equal to $S_W$.

The third observation is that the BH is not stable (in Minkowski space) because they can decay by emitting Hawking radiation.  States with large $S_W$ are long lived and can be treated  approximately as stable. States with a small $S_W$ decay on a time short compared to the classical horizon light crossing time and so cannot be treated even approximately as stable states. The inherent instability of states with small $S_W$ is in contrast to the situation in other bound systems where the low lying states, and in particular the ground state, are more stable than the highly excited states.

Additionally, states with small $S_W$ are ``quantum" (as opposed to semiclassical) and cannot be discussed reliably in the semiclassical approximation. This and the above mentioned instability prevent us from discussing in a reliable way the properties of the states with small values of $S_W$.

{\bf Acknowledgments:}
The research of RB and MH was supported by Israel Science Foundation grant no 239/10. MH research was supported by The Open University of Israel Research Fund. We thank Sunny Itzhaki for his help and advice, for many valuable and detailed discussions and for important and useful suggestions.
We also would like thank Joey Medved for enlightening discussions.

\end{document}